# Fluctuation electromagnetic conservative–dissipative interaction and heating of two closely spaced parallel plates in relative motion. Nonrelativistic approximation.2.

G.V. Dedkov, A.A. Kyasov

*Nanoscale Physics Group, Kabardino –Balkarian State University, Nalchik, 360004, Russian Federation*

*E-mail: gv_dedkov@mail.ru*

For the first time, we calculate the heating rate, attractive conservative and tangential dissipative fluctuation electromagnetic forces felt by a thick plate moving parallel to a closely spaced another plate in rest using the retarded nonrelativistic approximation of fluctuation electrodynamics. We argue that recently developed relativistic out of equilibrium theory of fluctuation electromagnetic interactions (Volokitin et. al., Phys.Rev. B78, 155437 (2008)) has serious drawbacks.

## 1.Introduction

This work goes a step further in comparison with our preceding paper [1], where we have calculated the fluctuation electromagnetic forces and heating rates in a system of two relatively moving thick featureless smooth plates divided by a vacuum gap of width $l$ using a non –retarded nonrelativistic approximation of fluctuation electrodynamics. We have got closed formulae for the force projections $F_x^{(2)}$, $F_z^{(2)}$ and heating rate $\dot{Q}^{(2)}$ corresponding to a plate with temperature $T_1$ moving with velocity $V$ relatively to another one, resting plate with temperature $T_2$. Superscript (2) denotes the proper configuration (see Fig.1(b)). Our basic configuration (Fig.1(a)) corresponds to a system being composed of a small spherical particle moving near a plate, and hereafter is denoted (1). The quantities $F_x^{(2)}$, $F_z^{(2)}$ and $\dot{Q}^{(2)}$ are given by [1]:

$$F_x^{(2)} = -\frac{\hbar S}{4\pi^3}\int_0^\infty d\omega \int_{-\infty}^{+\infty}dk_x \int_{-\infty}^{+\infty}dk_y k_x \frac{\exp(-2kl)}{\left|1-\exp(-2kl)\Delta_1(\omega^+)\Delta_2(\omega)\right|^2}\Delta_1''(\omega^+)\Delta_2''(\omega)\cdot$$
$$\cdot\left[\coth(\hbar\omega/2k_BT_2)-\coth(\hbar\omega^+/2k_BT_1)\right] \qquad (1)$$

$$F_z^{(2)} = -\frac{\hbar S}{4\pi^3}\int_0^\infty d\omega \int_{-\infty}^{+\infty} dk_x \int_{-\infty}^{+\infty} dk_y \, k \frac{\exp(-2kl)}{\left|1-\exp(-2kl)\Delta_1(\omega^+)\Delta_2(\omega)\right|^2} \cdot$$
$$\cdot \left[\Delta_1''(\omega^+)\Delta_2'(\omega)\coth(\hbar\omega^+/2k_B T_1) + \Delta_1'(\omega^+)\Delta_2''(\omega)\coth(\hbar\omega/2k_B T_2)\right] \quad (2)$$

$$\dot{Q}^{(2)} = \frac{\hbar S}{4\pi^3}\int_0^\infty d\omega \int_{-\infty}^{+\infty} dk_x \int_{-\infty}^{+\infty} dk_y \, \omega^+ \frac{\exp(-2kl)}{\left|1-\exp(-2kl)\Delta_1(\omega^+)\Delta_2(\omega)\right|^2} \Delta_1''(\omega^+)\Delta_2''(\omega) \cdot$$
$$\cdot \left[\coth(\hbar\omega/2k_B T_2) - \coth(\hbar\omega^+/2k_B T_1)\right] \quad (3)$$

$$\Delta_1(\omega) = \frac{\varepsilon_1(\omega)-1}{\varepsilon_1(\omega)+1}, \, \Delta_2(\omega) = \frac{\varepsilon_2(\omega)-1}{\varepsilon_2(\omega)+1}, \, \omega^+ = \omega + k_x V \quad (4)$$

where $\varepsilon_1(\omega)$ and $\varepsilon_2(\omega)$ are the dielectric permittivities of the plates, $S$ is the surface area, one time and two times primed quantities denote the corresponding real and imagine parts. As one directly sees from (1)-(3), the above formulae describe only electromagnetic field contributions related with surface evanescent modes $k > \omega/c$.

This paper aims to obtain more general retarded expressions in configuration 2 using the limit of small velocities, $\beta = V/c \to 0$, with $c$ being the speed of light in vacuum. Moreover, we restrict our consideration to the case of total thermal equilibrium. Similarly to [1], in order to get the necessary expressions we will use a "correspondence principle" between the configurations 1,2 (see Fig.1(a,b)). On this way, Eqs.(1)-(3) and our exact solution of the relativistic problem in configuration 1 [1-3] are the basic high lights being referred to.

## 2. Configuration 2: retarded interaction of parallel plates at a nonrelativistic relative velocity

A very important physical difference between the relativistic problem statements in configurations 1, 2 is that in the first one the presence of vacuum background is the basic standpoint and, correspondingly, we have only one large body (a thick plate) which can be in rest with respect to the background. A small particle, moving near the surface of resting plate, moves simultaneously with respect to the background. In this case, the resting plate may be or may not to be in thermal equilibrium with the background radiation. This condition directly determines the structure of fluctuating electromagnetic field near the plate. For configuration 2, in contrast, the problem statement in dynamic situation needs to be more elaborate even at $T_1 = T_2 = T_3 = T$, because only one of the plates can be in rest respectively to the background, whereas another plate will be braking due to the interaction with background. This fact has been completely ignored in the proposed theory of fluctuation dissipative forces and heat exchange [4-6], developed to date. On the other hand, several attempts of other authors to develop a relativistic method for calculating the dissipative force $F_x^{(2)}$ [7-10] have resulted in zero dissipative (friction) force in the limit $c \to \infty$, that is in principal conflict with the well established results of the norelativistic consideration [11,12].

Quite recently, the presence of vacuum background in configuration 2 has been discussed in relation with thermal Casimir forces, $F_z^{(1)}$, $F_z^{(2)}$, which have been calculated out of thermal equilibrium is the static case $V = 0$ [13,14]. Incidentally, the resulting formula for $F_z^{(2)}$ which follows from [6] in this case, turns out to disagree with [13,14]. And what is more, as we have shown in [1], even the corresponding norelativistic expressions for the quantities $F_z^{(2)}$ and $\dot{Q}^{(2)}$ which stem from [6], prove to be in error. Therefore, a further elaboration of the dynamic problem of the fluctuation electromagnetic interaction in configuration 2 seems to be of crucial importance.

Bearing in mind a "correspondence principle" between the configurations 1,2 [1] and using our exact relativistic expressions for $F_x^{(1)}$, $F_z^{(1)}$ and $\dot{Q}^{(1)}$ in configuration 1 at $T_1 = T_2 = T_3 = T$ [1], we obtain

$$F_z^{(2)}(l) = -\frac{\hbar S}{4\pi^3} \int_0^\infty d\omega \int_{-\infty}^{+\infty} dk_x \int_{-\infty}^{+\infty} dk_y \cdot$$

$$\cdot \left[ \begin{pmatrix} \dfrac{\operatorname{Im}\Delta_{1e}(\omega^+)\operatorname{Re}(q_0 \exp(-2q_0 l)\Delta_{2e}(\omega))}{\left|1-\exp(-2q_0 l)\Delta_{1e}(\omega^+)\Delta_{2e}(\omega)\right|^2}\coth\left(\dfrac{\hbar\omega}{2k_B T}\right) + \\ + \dfrac{\operatorname{Im}\Delta_{1m}(\omega^+)\operatorname{Re}(q_0 \exp(-2q_0 l)\Delta_{2m}(\omega))}{\left|1-\exp(-2q_0 l)\Delta_{1m}(\omega^+)\Delta_{2m}(\omega)\right|^2}\coth\left(\dfrac{\hbar\omega}{2k_B T}\right) \end{pmatrix} + \\ + \begin{pmatrix} \dfrac{\operatorname{Re}\Delta_{1e}(\omega^+)\operatorname{Im}(q_0 \exp(-2q_0 l)\Delta_{2e}(\omega))}{\left|1-\exp(-2q_0 l)\Delta_{1e}(\omega^+)\Delta_{2e}(\omega)\right|^2}\coth\left(\dfrac{\hbar\omega}{2k_B T}\right) + \\ + \dfrac{\operatorname{Re}\Delta_{1m}(\omega^+)\operatorname{Im}(q_0 \exp(-2q_0 l)\Delta_{2m}(\omega))}{\left|1-\exp(-2q_0 l)\Delta_{1m}(\omega^+)\Delta_{2m}(\omega)\right|^2}\coth\left(\dfrac{\hbar\omega}{2k_B T}\right) \end{pmatrix} \right] \quad (5)$$

$$F_x^{(2)}(l) = -\frac{\hbar S}{4\pi^3} \int_0^\infty d\omega \int_{-\infty}^{+\infty} dk_x \int_{-\infty}^{+\infty} dk_y k_x \cdot$$

$$\cdot \left[ \frac{\operatorname{Im}\Delta_{1e}(\omega^+)\operatorname{Im}(\exp(-2q_0 l)\Delta_{2e}(\omega))}{\left|1-\exp(-2q_0 l)\Delta_{1e}(\omega^+)\Delta_{2e}(\omega)\right|^2} + \frac{\operatorname{Im}\Delta_{1m}(\omega^+)\operatorname{Im}(\exp(-2q_0 l)\Delta_{2m}(\omega))}{\left|1-\exp(-2q_0 l)\Delta_{1m}(\omega^+)\Delta_{2m}(\omega)\right|^2} \right] \cdot \quad (6)$$

$$\cdot \left[ \coth\left(\frac{\hbar\omega}{2k_B T}\right) - \coth\left(\frac{\hbar\omega^+}{2k_B T}\right) \right]$$

$$\dot{Q}^{(2)}(l) = \frac{\hbar S}{4\pi^3} \int_0^\infty d\omega \int_{-\infty}^{+\infty} dk_x \int_{-\infty}^{+\infty} dk_y \omega^+ \cdot$$

$$\cdot \left[ \frac{\operatorname{Im}\Delta_{1e}(\omega^+)\operatorname{Im}(\exp(-2q_0 l)\Delta_{2e}(\omega))}{\left|1-\exp(-2q_0 l)\Delta_{1e}(\omega^+)\Delta_{2e}(\omega)\right|^2} + \frac{\operatorname{Im}\Delta_{1m}(\omega^+)\operatorname{Im}(\exp(-2q_0 l)\Delta_{2m}(\omega))}{\left|1-\exp(-2q_0 l)\Delta_{1m}(\omega^+)\Delta_{2m}(\omega)\right|^2} \right] \cdot \quad (7)$$

$$\cdot \left[ \coth\left(\frac{\hbar\omega}{2k_B T}\right) - \coth\left(\frac{\hbar\omega^+}{2k_B T}\right) \right]$$

$$\Delta_{ie}(\omega) = \frac{q_0 \varepsilon_i(\omega) - q_i}{q_0 \varepsilon(\omega) + q_i}, \quad \Delta_{im}(\omega) = \frac{q_0 \mu_i(\omega) - q_i}{q_0 \mu_i(\omega) + q_i}, \quad i=1,2 \quad (8)$$

$$q_0 = (k^2 - \omega^2/c^2)^{1/2}, \quad k^2 = |\mathbf{k}|^2 = k_x^2 + k_y^2, \quad q_i = \left(k^2 - (\omega^2/c^2)\varepsilon_i(\omega)\mu_i(\omega)\right)^{1/2} \quad (9)$$

where $i=1,2$ denotes the plates, $\varepsilon_i(\omega)$ and $\mu_i(\omega)$ are the involved dielectric permittivities and magnetic permeabilities of the materials. In notations of other authors, [4,5,6,7,8,9,10,11], $\Delta_{ie}(\omega)$ and $\Delta_{im}(\omega)$ correspond to the reflection amplitudes of the electromagnetic waves with $P-$ and $S-$polarization. Eqs.(4) evidently describe the non –retarded limits of $\Delta_{ie}(\omega)$, while the proper limits of $\Delta_{im}(\omega)$ prove to be zero. In this case and at thermal equilibrium, a total correspondence between (1)-(3) and (5)-(7) is quite obvious. However, in relation with (5)-(7), two points need to be clarified.

First, the obtained formulae do not take into account a contribution from vacuum background in direct form, that is well justified at $\beta \to 0$. Indirectly, the presence of vacuum background displays as an asymmetry between the first and second plates. That seems to be completely obvious if one takes into account that the second plate is in rest respectively to the background, while the first one is moving.

Second, as the integration limits over the wave –vector projections $k_x, k_y$ are infinite, $(-\infty,\infty)$, formulae (5)-(7) describe both contributions of surface evanescent modes ($k > \omega/c$), and of surface propagating modes ($k < \omega/c$). The equilibrium condition $T_1 = T_2 = T_3 = T$ automatically ensures the lack of "radiation wind" terms being proportional to the absorption coefficients $(1-|\Delta_{ie}|^2)$, $(1-|\Delta_{im}|^2)$, while the structure of fluctuation electromagnetic field in this case manifests an oscillating character at $k < \omega/c$ [15,16]. If a system is out of thermal equilibrium, the contribution of surface propagating modes, in general, does not merely have an oscillating structure, whereas the contribution of evanescent modes does not change. Therefore, formulae (5)-(7) at $k > \omega/c$ are valid at arbitrary temperatures $T_1, T_2, T_3$.

So, despite that formulae (5)-(7) do not solve general relativistic problem out of thermal equilibrium in configuration 2, they may be considered as the first step on the way of development of the general relativistic theory. Now, let us demonstrate that formulae (5)-(7) are in accordance with the known general solution of the problem in configuration 1 [1,2,3], if use is made of the limit of rarified medium for the moving plate:

$\varepsilon_1(\omega) - 1 = 4\pi n_1 \alpha_e(\omega) \to 0$, $\mu_1(\omega) - 1 = 4\pi n_1 \alpha_m(\omega) \to 0$, where $n_1$ is the atomic density of the first plate, $\alpha_{e,m}(\omega)$ are the involved dielectric and magnetic polarizabilities. Then, with account of the relations

$$\Delta_{1e}(\omega) \cong \frac{\pi n_1}{q_0^2}\left[\alpha_e(\omega)(2k^2 - \omega^2/c^2) + \alpha_m(\omega)\omega^2/c^2\right] \qquad (10)$$

$$\Delta_{1m}(\omega) \cong \frac{\pi n_1}{q_0^2}\left[\alpha_m(\omega)(2k^2 - \omega^2/c^2) + \alpha_e(\omega)\omega^2/c^2\right] \qquad (11)$$

$$F_z^{(1)}(z) = -\frac{1}{n_1 S}\frac{dF_z^{(2)}(l)}{dl}\bigg|l = z \qquad (12)$$

$$F_x^{(1)}(z) = -\frac{1}{n_1 S}\frac{dF_x^{(2)}(l)}{dl}\bigg|l = z \qquad (13)$$

$$\dot{Q}^{(1)}(z) = -\frac{1}{n_1 S}\frac{d\dot{Q}^{(2)}(l)}{dl}\bigg|l = z \qquad (14)$$

we get

$$F_z^{(1)}(z) = -\frac{\hbar}{2\pi^2}\int_0^\infty d\omega \int_{-\infty}^{+\infty} dk_x \int_{-\infty}^{+\infty} dk_y \cdot$$
$$\cdot\left\{\begin{array}{l}\alpha_e''(\omega^+)\mathrm{Re}[\exp(-2q_0 z)R_e(\omega,\mathbf{k})]\coth\left(\frac{\hbar\omega^+}{2k_B T}\right) + \\ + \alpha_e'(\omega^+)\mathrm{Im}[\exp(-2q_0 z)R_e(\omega,\mathbf{k})]\coth\left(\frac{\hbar\omega}{2k_B T}\right) + (e \leftrightarrow m)\end{array}\right\} \qquad (15)$$

$$F_x^{(1)}(z) = -\frac{\hbar}{2\pi^2} \int_0^\infty d\omega \int_{-\infty}^{+\infty} dk_x \int_{-\infty}^{+\infty} dk_y k_x \cdot \alpha_e''(\omega^+) \cdot \left\{ \begin{array}{l} \mathrm{Im}\left[\frac{\exp(-2q_0 z)}{q_0} R_e(\omega, \mathbf{k})\right] \cdot \\ \cdot \left[\coth\left(\frac{\hbar\omega}{2k_B T}\right) - \coth\left(\frac{\hbar\omega^+}{2k_B T}\right)\right] \end{array} + (e \to m) \right\} \quad (16)$$

$$\frac{dQ^{(1)}(z)}{dt} = \frac{\hbar}{2\pi^2} \int_0^\infty d\omega \int_{-\infty}^{+\infty} dk_x \int_{-\infty}^{+\infty} dk_y \omega^+ \alpha_e''(\omega^+) \cdot \left\{ \begin{array}{l} \mathrm{Im}\left[\frac{\exp(-2q_0 z)}{q_0} R_e(\omega, \mathbf{k})\right] \cdot \\ \cdot \left[\coth\left(\frac{\hbar\omega}{2k_B T}\right) - \coth\left(\frac{\hbar\omega^+}{2k_B T}\right)\right] \end{array} + (e \to m) \right\} \quad (17)$$

$$R_e(\omega, \mathbf{k}) = \Delta_{2e}(\omega)(2k^2 - \omega^2/c^2) + \Delta_{2m}(\omega)\omega^2/c^2 \quad (18)$$

where the terms $(e \leftrightarrow m)$ are defined by the same expressions when replacing subscripts "e" by "m". Besides, the subscripts "2" in (18) remind that material properties correspond to the second (resting) plate. Eqs.(16)-18) are in complete agreement with their relativistic analogs presented in [1,2,3] and follow from those ones at $\beta = V/c \to 0$, $\gamma = (1-\beta^2)^{-1/2} \to 1$, $T_1 = T_2 = T_3 = T$.

**3. Discussion**

In our preceding paper [1] we have shown that the recently developed relativistic theory in configuration 2, [6], has serious drawbacks in the nonrelativistic limit. In view of the results obtained in the previous section, that becomes more clear with account of retardation effects. Thus, the expression for $F_z^{(1)}(z)$ in Ref. [6] (Eq.(28) at $\beta \to 0$, in our notations) can be written in the form

$$F_z^{(1)}(z) = -\frac{\hbar}{2\pi^2} \mathrm{Im} \int_0^\infty d\omega \int_{-\infty}^{+\infty} dk_x \int_{-\infty}^{+\infty} dk_y \cdot \exp(-2q_0 z)(k^2 - \omega^2/c^2) \cdot$$
$$\cdot [\alpha_e(\omega^+)\Delta_e(\omega) + \alpha_m(\omega^+)\Delta_m(\omega)] \cdot \left[\coth\left(\frac{\hbar\omega^+}{2k_B T_1}\right) + \coth\left(\frac{\hbar\omega}{2k_B T_2}\right)\right] \quad (19)$$

Comparison between (19) taken at $T_1 = T_2 = T$ and (15) seems to be very instructive. First, Eq.(19), as well as its analog in configuration 2 in the non–retarded limit, has completely incorrect dependence on temperatures (see also [1]). However, this error is not the last one. Consider, for instance, a simpler case $\alpha_m(\omega) = 0$, $V = 0$. Then Eqs.(19) and (15) take the form

$$F_z^{(1)}(z) = -\frac{\hbar}{\pi^2} \text{Im} \int_0^\infty d\omega \int_{-\infty}^{+\infty} dk_x \int_{-\infty}^{+\infty} dk_y \exp(-2q_0 z)(k^2 - \omega^2/c^2)\alpha_e(\omega)\Delta_e(\omega)\coth\left(\frac{\hbar\omega}{2k_B T}\right) \quad (20)$$

$$F_z^{(1)}(z) = -\frac{\hbar}{2\pi^2} \text{Im} \int_0^\infty d\omega \int_{-\infty}^{+\infty} dk_x \int_{-\infty}^{+\infty} dk_y \cdot \exp(-2q_0 z)\alpha_e(\omega) \cdot$$
$$\cdot \left[\Delta_e(\omega)(2k^2 - \omega^2/c^2) + \Delta_m(\omega)\omega^2/c^2\right] \cdot \coth\left(\frac{\hbar\omega}{2k_B T}\right) \quad (21)$$

As a matter of fact, formula (20) does not contain a contribution with $\Delta_m(\omega)$, whereas the coefficient $(k^2 - \omega^2/c^2)$ differs from $(2k^2 - \omega^2/c^2)$ in Eq.(21). This contradicts to the well recognized theory of the Casimir–Polder force [17,18], because Eq.(20) proves to be independent of the reflection coefficient of S-polarized electromagnetic modes. Contrary to that, formula (21) turns out to be in full agreement with the theory [17,18] and, as we have shown in Section 3, it follows from configuration 2 in the limit of rarified body. In addition to this, in Refs. [4,5], as far as the dissipative force $F_x^{(1)}(z)$ and heating rate $\dot{Q}^{(1)}$ is concerned, we claim that the factors, identical to ours $R_{e,m}(\omega, \mathbf{k})$ (see (18)) are also in error.

Now, and let us compare Eq.(6) with its counterpart, Eq.(22) in Ref. [6]. It is worth noticing that the thermal state of the vacuum background is not defined by the authors. In the retarded limit at $\beta \to 0$, $T_1 = T_2 = T$, in our notations, Eq.(22) in [6] takes the form

$$F_x^{(2)}(l) = -\frac{\hbar S}{16\pi^3} \int_0^\infty d\omega \int_{k<\omega/c} d^2k \cdot k_x$$
$$\cdot \left[\frac{\left(1-|\Delta_{1e}(\omega^+)|^2\right)\left(1-|\Delta_{2e}(\omega)|^2\right)}{\left|1-\exp(2i|q_0|l)\Delta_{1e}(\omega^+)\Delta_{2e}(\omega)\right|^2} + (e \leftrightarrow m)\right] \cdot \left[\coth\left(\frac{\hbar\omega}{2k_B T}\right) - \coth\left(\frac{\hbar\omega^+}{2k_B T}\right)\right] -$$
$$-\frac{\hbar S}{4\pi^3} \int_0^\infty d\omega \int_{k>\omega/c} d^2k \cdot k_x \exp(-2q_0 l) \cdot$$
$$\cdot \left[\frac{\text{Im}\Delta_{1e}(\omega^+)\text{Im}(\Delta_{2e}(\omega))}{\left|1-\exp(2i|q_0|l)\Delta_{1e}(\omega^+)\Delta_{2e}(\omega)\right|^2} + (e \leftrightarrow m)\right] \cdot \left[\coth\left(\frac{\hbar\omega}{2k_B T}\right) - \coth\left(\frac{\hbar\omega^+}{2k_B T}\right)\right] \quad (22)$$

Unlike [6], in writing (22) we have assumed, for uniformity, that the moving plate is the first one (as shown in Fig.1(b)). Comparison of (6) and (22) shows that the terms corresponding to surface evanescent modes ($k > \omega/c$) coincide with each other, whereas the terms corresponding to surface propagating modes, are essentially different. Eq.(6) does not contain the absorption coefficients $\left(1-|\Delta_{1e}(\omega^+)|^2\right)$, $\left(1-|\Delta_{2e}(\omega)|^2\right)$, etc., because the system is assumed to be in total thermal equilibrium, $T_1 = T_2 = T_3 = T$, while the equilibrium fluctuation electromagnetic field proves to have the structure of oscillating, standing wave. Besides, the presence of relative motion does not principally violate state of the field seen in reference frame of the moving plate: $\Delta_{e,m}(\omega) \to \Delta_{e,m}(\omega^+)$. Therefore, correct statement of the problem in configuration 2 at $T_1 = T_2 = T_3 = T$ could not result in the radiation wind terms. These terms may appear out of equilibrium, but this takes us off the limits of the equilibrium theory of Lifshitz [19]. And what is more, if the wind terms appear in the tangential force $F_x^{(2)}$, so they do in $F_z^{(2)}$, as well. In the last case, at $V = 0$, that has been clearly demonstrated in [18]. However, in [6] the proper contributions are absent, whereas the wind terms make contribution into $F_x^{(2)}$ even under the equilibrium conditions (see (22) and Ref. [4]).

In fact, the authors of [4,5,6], starting from their first papers on fluctuation dissipative forces, [4], have proceeded just from the equilibrium Lifshitz theory [19], trying to adjust a classical solution to the dynamic and non–equilibrium situations. Therefore, the relativistic out of equilibrium theory [6], as well as the previous nonrelativistic equilibrium [4,5] and non–equilibrium (like (22)) modifications are in error.

**4. Conclusion**

With account of our exact solution to the relativistic problem of fluctuation electromagnetic interaction in configuration 1 (a small particle moving near a wall), and using a correspondence principle between configuration 1 and 2 (two thick featureless parallel plates in relative motion), we have obtained the self consistent formulae for the conservative –dissipative forces and heating rate in configuration 2 in the retarded nonrelativistic approximation of fluctuation electrodynamics. Closed transition rules between the involved configurations are formulated. The obtained formulae may be regarded as important referring ones when solving the general relativistic problem in configuration 2. However, the resulting expressions disagree with the theory of Volokitin et. al. [4,5,6] in some principal points. We argue that basics of the theory [4,5,6] is not well defined, while the resulting formulae have numerous errors.

**References**

[1] G.V.Dedkov, A.A.Kyasov, arXiv: 0904.0124v1 [cond-mat.other]
[2] G.V.Dedkov, A.A.Kyasov, J.Phys.: Condens. Matter 20, 354006 (2008)
[3] G.V.Dedkov, A.A.Kyasov, Phys. Solid State 51(1), 1 (2009)

[4] A.I.Volokitin, B.N.J.Persson, Phys. Low.-Dim/ Struct. 7/8 (1998); J.Phys.:Condens. Matter 11,345 (1999); Phys. Rev.B63, 205404 (2001); Phys. Rev. B65, 115419(2001)

[5] A.I. Volokitin and B. N. Persson, Usp. Fiz. Nauk 177 (9), 921 (2007) [Phys—Usp. 50 (9), 879 (2007)]; Rev. Mod. Phys. 79, 1291 (2007)

[6] A.I.Volokitin, B.N.J. Persson, Phys. Rev. B78, 155437 (2008); arXiv: 0807.1004v1 [cond-mat.other] 7 Jul 2008

[7] L.S.Levitov, Eur. Phys. Lett. 8, 488 (1989)

[8] V.G.Polevoi, Sov.Phys. JETP 71(6)1119 (1990)

[9] V.E.Mkrtchian, Phys. Lett. A207, 299(1995)

[10] I.Dorofeyev, H.Fuchs, B.Gotsmann, and J.Jersch, Phys. Rev. B64, 035403 (2001)

[11] M.S.Tomassone and A.Widom, Phys. Rev. B56, 493 (1997)

[12] J.B.Pendry, J.Phys.: Condens. Matter 9, 10301 (1997)

[13] M.Antezza, L.P.Pitaevskii, and S.Stringari, Phys. Rev. Lett. 95, 113202 (2005)

[14] M.Antezza, L.P.Pitaevskii, S.Stringari, and V.B.Svetovoy, arXiv: 0706.1860v2 [cond-mat.stat-mech] 5Feb 2008

[15] K.Joulain, R.Carminati, J.-J.Greffet, Phys. Rev. B68, 245405 (2003)

[16] G.V.Dedkov, A.A.Kyasov, Tech. Phys. Lett.32(3), 223(2006)

[17] M.Antezza, L.P.Pitaevskii, and S.Stringari, Phys. Rev. A70, 053619(2004)

[18] V.B.Bezerra, G.L.Klimchitskaya, V.M.Mostepanenko, and C.Romero, arXiv:0809.5229v1 [quant-ph] 30 Sep 2008

[19] E.M.Lifshitz, Sov.Phys. JETP 2, 73(1956)

FIGURES:

FIGURE 1(a)

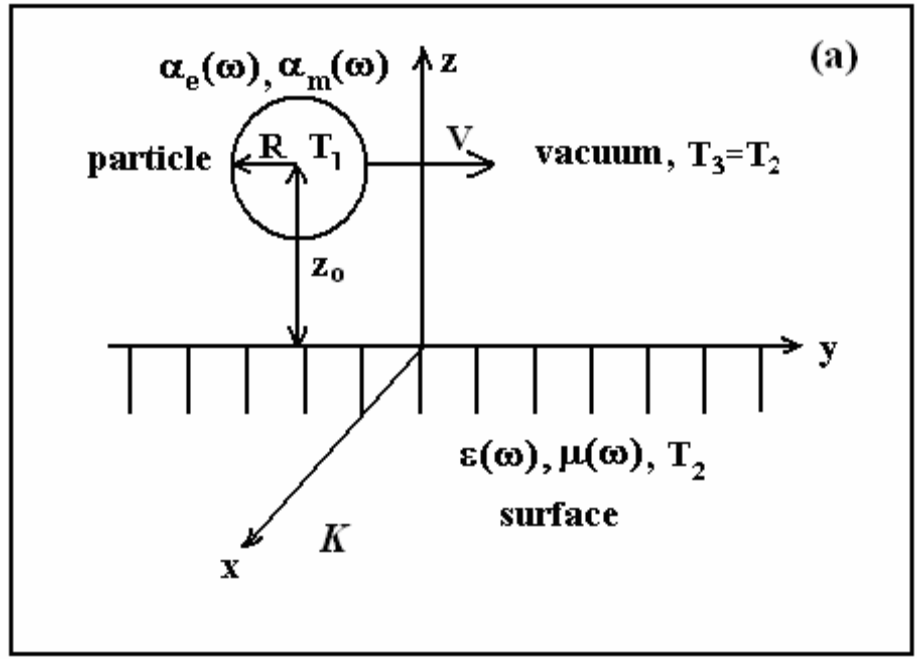

Fig.1(a) Configuration 1. Geometry of motion of a particle and a Cartesian reference frame associated with the surface of the medium (system $K$). The Cartesian axes ($x', y', z'$) of the particle rest frame $K'$ are not shown.

FIGURE 1(b)

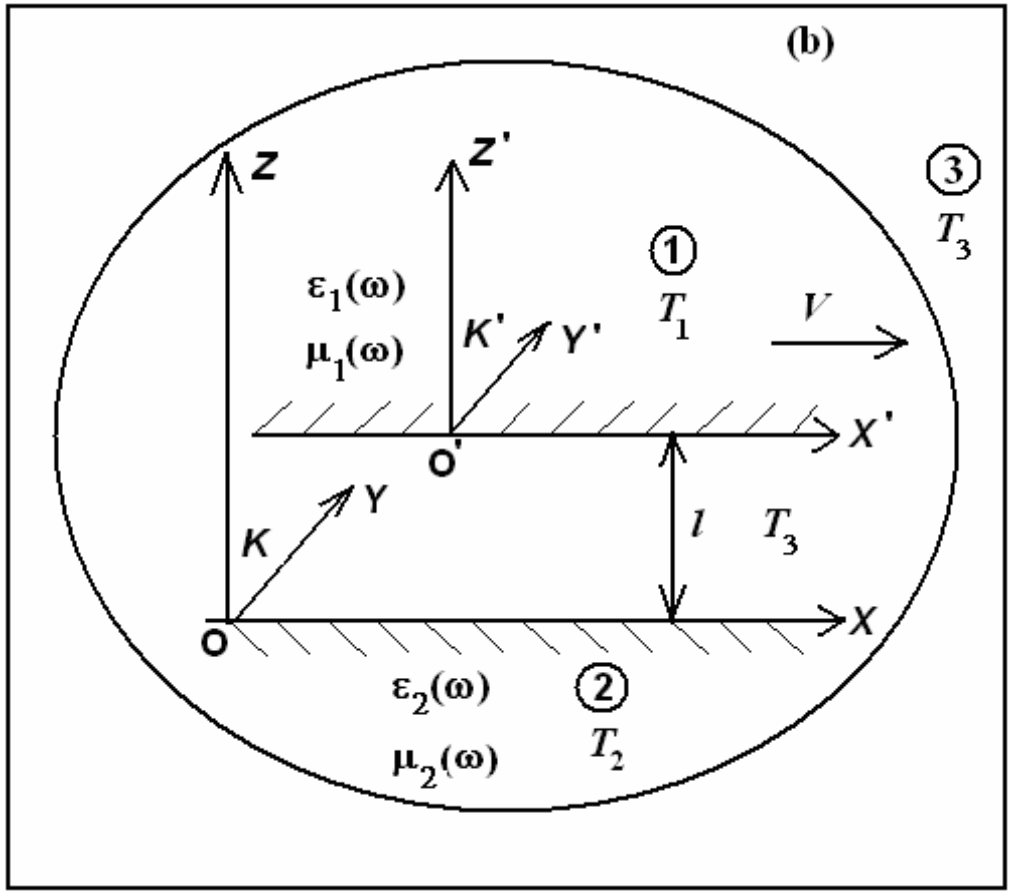

Fig.1(b) Configuration 2, corresponding to large thick plates 1 and 2 with temperatures $T_1$ and $T_2$ in the rest frame of each one, respectively. $K$ and $K'$ are the corresponding Cartesian reference frames. Surrounding vacuum background, in general, may have the temperature $T_3$.